\documentclass[journal,twocolumn]{IEEEtran}

\usepackage{graphicx, extarrows, pdfsync, url}
\usepackage{epstopdf, enumerate, color, multirow}
\usepackage[center]{caption}

\newcommand{\mygraphic}[1]{\includegraphics[height=#1]{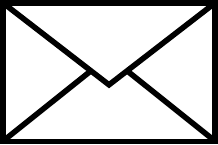}}
\newcommand{\myenv}{(\raisebox{0pt}{\mygraphic{.6em}})}
\newcommand{\myauthor}[1]{#1~\myenv}

\begin{document}

\title{Identifying ECUs Using Inimitable Characteristics of Signals in Controller Area Networks}

\author{Wonsuk Choi,
        Hyo Jin Jo,
        Samuel Woo,
        Ji Young Chun,
        Jooyoung Park,
        and~\myauthor{Dong Hoon Lee}%,~\IEEEmembership{Member,~IEEE}% <-this % stops a space

%\thanks{Manuscript received XXX, XX, 2015; revised XXX, XX, 2015.}% <-this % stops a space
%\thanks{Wonsuk Choi, Hyo Jin Jo, Samuel Woo, Ji Young Chun, and Dong Hoon Lee are with the Graduate School of Information Security, Korea University, Seoul, Republic of Korea (e-mail: wonsuk85.choi@gmail.com; hyojinjo86@gmail.com; samuelwoo@korea.ac.kr; jychun@korea.ac.kr; donghlee@korea.ac.kr).
%Jooyoung Park is with the Department of Control and Instrumentation Engineering, Korea University, Sejong City, Republic of Korea (e-mail: parkj@korea.ac.kr) }
}

%\markboth{IEEE Transactions on Vehicular Technology,~Vol.~XX, No.~XX, XXX~2015}
{}
%{Shell \MakeLowercase{\textit{et al.}}: Bare Demo of IEEEtran.cls for Journals}

\maketitle

\begin{abstract}
\boldmath
In the last several decades, the automotive industry has come to incorporate the latest Information and Communications (ICT) technology, increasingly replacing mechanical components of vehicles with electronic components. These electronic control units (ECUs) communicate with each other in an in-vehicle network that makes the vehicle both safer and easier to drive. Controller Area Networks (CANs) are the current standard for such high quality in-vehicle communication. Unfortunately, however, CANs do not currently offer protection against security attacks. In particular, they do not allow for message authentication and hence are open to attacks that replay ECU messages for malicious purposes. Applying the classic cryptographic method of message authentication code (MAC) is not feasible since the CAN data frame is not long enough to include a sufficiently long MAC to provide effective authentication. In this paper, we propose a novel identification method, which works in the physical layer of an in-vehicle CAN network. Our method identifies ECUs using inimitable characteristics of signals enabling detection of a compromised or alien ECU being used in a replay attack. Unlike previous attempts to address security issues in the in-vehicle CAN network, our method works by simply adding a monitoring unit to the existing network, making it deployable in current systems and compliant with required CAN standards. Our experimental results show that the bit string and classification algorithm that we utilized yielded more accurate identification of compromised ECUs than any other method proposed to date. The false positive rate is more than 2 times lower than the method proposed by P.-S. Murvay et al. This paper is also the first to identify potential attack models that systems should be able to detect.
\end{abstract}


% \fi
%%\IEEEpeerreviewmaketitle

\section{Introduction}
In recent years, the automotive industry has increasingly replaced mechanical components of vehicles with electronic components, using the latest Information and Communications (ICT) technology.
Electronic Control Units (ECUs) were originally proposed to ensure optimal engine performance, particularly in terms of efficient gasoline and oil consumption \cite{kassakian1996automotive}.
More recently, automotive manufactures have installed ECUs not only for engine control but also to perform various functions for driver safety and convenience such as Anti-lock Brake Systems and Intelligent Parking Assist Systems.
Luxury sedans now contain 50-70 independent ECUs \cite{ChasharkI}.
These ECUs form a network to communicate with each other that is divided into several sub-networks.
For example, ECUs for chassis control are connected with the in-vehicle Controller Area Network (CAN).
The CAN was designed by Robert Bosch GimbH in 1983 for automotive applications to provide reliable in-vehicle communication between ECUs \cite{CAN_Standard}.
The CAN's reliability and simple network structure has made it the standard for communication as an in-vehicle network for over 30 years.

Despite the advantages of the CAN, it was not designed with security features in mind; hence, the network is open to attack from malicious messages intended to cause malfunctions.
In fact, previous research has analyzed the security features of modern vehicles and found that the fundamental reason why the CAN is subject to attack is that it does not support message authentication \cite{CarsharkII,ChasharkI}.
Unfortunately, the existing method for message authentication cannot be applied to the in-vehicle CAN network: it is difficult to use message authentication code (MAC) because there is no field for message authentication in the CAN data frame.
Even if MAC could be transmitted through the CAN data field, the effectiveness of the network to authenticate messages would be extremely limited because the length of the CAN data field is only 1-8 bytes while the MAC needed for adequate security is more than 20 bytes.

To address this vulnerability, some research has been conducted on message authentication in the in-vehicle CAN network \cite{Tesla,Cyber_security_for_CAN,CANAuth}.
However, previous work did not succeed in solving the fundamental inadequacy of the current CAN to incorporate security features: the methods of previous researchers simply cannot be directly applied into the current CAN system. Herrewege et al. \cite{CANAuth} developed a CAN+ protocol initially proposed by Ziermann et al. \cite{ziermann2009can+}, which inserts extra bits between the sampling points of a CAN bus interface; however, in order to use this CAN+ protocol, CAN transceivers would have to have a higher bit rate than they currently do, meaning that all ECUs would have to be replaced.
And, both Groza et al. \cite{Tesla} and Lin et al. \cite{Cyber_security_for_CAN} used methods that send additional messages for message authentication, which leads to a rapid increase in the bus load (to more than 50$\%$). In general, the bus load has to remain under 50$\%$ of the maximum in order to preserve a stable communication environment \cite{Woo}.
In other words, this method would also require replacing existing devices with new ones with higher performance capabilities in order to maintain stable communication.
Prior proposals therefore require complete overhauls of the current and ubiquitous CAN system and for that reason do not offer practical solutions.

In this paper, we propose a novel identification method for the in-vehicle CAN network that can be directly applied into the current system.
Our method utilizes an add-on to the current in-vehicle CAN network, without requiring replacement of current ECUs, and complies with the standard CAN in use today.
We build on the idea of source identification using signal characteristics in CAN ECUs, as first suggested by Murvay et al. \cite{Cosic}.
However, this work reflected two weaknesses. First, Murvay et al. did not evaluate their method on ECUs that manage critical functions using the high-speed CAN network, but only used bit rates used in ECUs that manage simpler functions on the low-speed CAN network.
Since it is much easier to achieve device identification using the lower bit rate \cite{Wired_physical_layer}, they did not appropriately simulate the CAN environment. Our method overcomes this weakness by using the same bit rate that is used on the high-speed CAN when ECUs perform critical functions (i.e, 500K b/s). This is a significantly higher bit rate than Murvay et al. applied \cite{Cosic}.
Second, Murvay et al. failed to consider how situations involving collisions would impact signal identification using, as he did, the identifier field. A collision occurs when more than two ECUs send data at the same time. In a collision situation, signals are generated from multiple ECUs, an arbitration system assigns priorities to the signals, and lower priority signals are cut off. The identifier field, which Murvay et al. relied upon, is unable to differentiate signals under these collision circumstances, and therefore is insufficient to accurately identify signal characteristics.
By contrast, our method measures signals using the extended identifier field in the extended frame format of CAN, which is able to accurately identify signals from all ECUs free of any collision confusion that may be present in the identifier field.
In addition to accurately measuring all signals by using the extended identifier field instead of the identifier field and simulating the bit rate of the high-speed CAN, our method is more than twice as accurate as previous research in identifying which ECU sent a given message, in terms of false positive rate.

\subsection{Our Contributions}
We propose our method for ECU identification based on the inimitable characteristics of signals in the physical layer. Our method is simpler and more transparent than ones using message authentication, which in any case are not feasible in existing in-vehicle CAN networks. Even if a message authentication method could be developed compatible with existing CANs, it would require the additional cost of a key exchange or other key management system. Our method is able to identify whether ECUs are valid or malicious without relying on such method authentication systems. Our identification method is also superior to the previous approach of Murvay et al. \cite{Cosic} in terms of accuracy. Our main contributions are as follows:
	\begin{itemize}
            \item Our method is able to identify ECUs only by installing an additional device, meaning that our method can be directly applied into current vehicles.
            \item Our method complies with current CAN standards and so does not require replacement or alteration of ECUs that have been installed in in-vehicle networks.
            \item Our method improves upon the work of Murvay et al. \cite{Cosic} in that it analyzes more features of in-vehicle CANs, including those that control critical functions; identifying ECUs accurately even under collision conditions when more than two ECUs transmit data simultaneously; and testing our method at the same bit rate (500K b/s) as the bit rate of the in-vehicle CAN network..
     \end{itemize}
The rest of the paper is structured as follows: Section \ref{Motivations} describes our motivation for this work to address limitations of the in-vehicle CAN network. In Section \ref{Backgrounds}, we describe background needed to understand our method. In Sections \ref{System model} and 5, we propose our system model and method for ECU identification, respectively. In Section 6, we present our experimental design and results, showing the accuracy of our method. Section 7 describes related work revealing research trends. Finally, we describe future work and our conclusion in Sections 8 and 9, respectively.

%%%%%%%%%%%%%%%%%%%%%%%%%%%%%%%%%%%%%%%%%%%%%%%%%%%%%%%%%%%%%%%%%%%%%%%%%%%%%%%%%%%%%%%%%%%%%%%%%%%%%%%%
\begin{figure*}[t]
\centering
\includegraphics[width=7in]{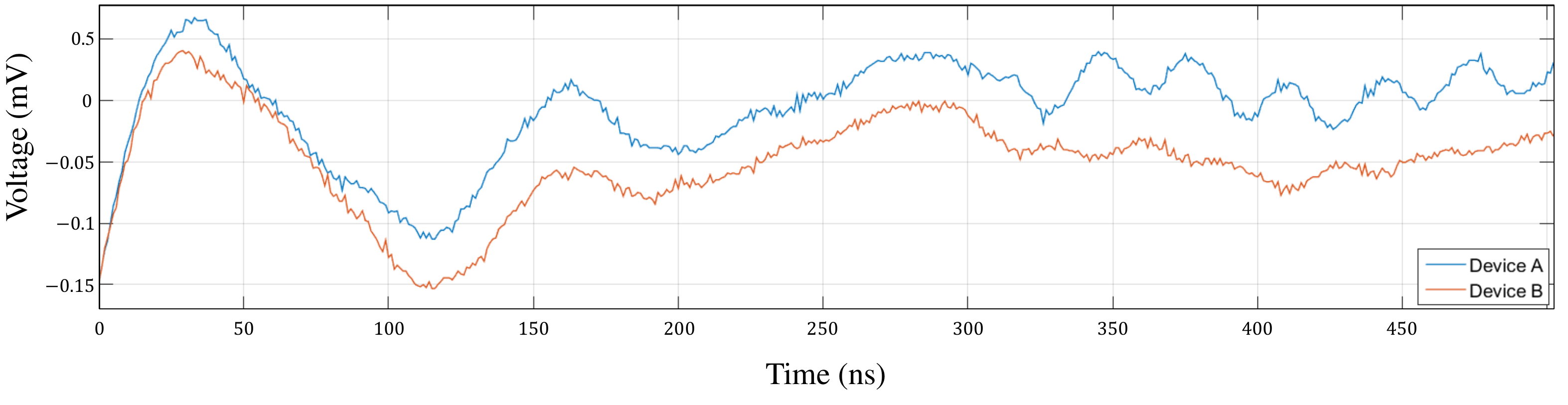}
\caption{An example of variations in signaling}
\label{An example of variations in signaling}
\end{figure*}
%%%%%%%%%%%%%%%%%%%%%%%%%%%%%%%%%%%%%%%%%%%%%%%%%%%%%%%%%%%%%%%%%%%%%%%%%%%%%%%%%%%%%%%%%%%%%%%%%%%%%%%%
\section{Motivations}
\label{Motivations}
Most security features are not compatible with the CAN, so that it has been especially challenging to provide message authentication to effectively protect the CAN from malicious attackers.  Without effective message authentication, every ECU in the in-vehicle CAN network is able to freely communicate with every other ECU, opening the system up to replay attacks. Since CAN packets contain no authenticator field, any ECU can impersonate any other ECU in the in-vehicle CAN network.

The lack of authentication makes the CAN vulnerable to a variety of potential attacks, compromising reliable functioning of vehicles.
Koscher et al. were the first to demonstrate several examples of attacks on modern vehicles \cite{ChasharkI}.
It is possible for an adversary to have the ability to systematically control a wide array of components including the engine, instrument panel, radio, heating and cooling, lights, brakes, and locks.
If an adversary attacks a vehicle during operation of the vehicle, the lives of the driver and passengers can be in danger.
For example, Koscher et al. explained that an adversary could forcibly and completely disengage the brakes.
Koscher et al. were only able to test potential threats using direct physical access with a laptop connected to internal vehicle networks.
Next, Checkoway et al. demonstrated long-range or indirect physical access attacks \cite{CarsharkII}.
Applying their adversary model, they found that it would be difficult for vehicle owners to notice attacks.
They connected in-vehicle networks to external networks such as Bluetooth and audio CD devices. For example, they suggested that an adversary might deliver malicious input by encoding it as a song file.
When the modified media file is played on car audio system, malicious CAN packets causing intentional malfunctions are transmitted into in-vehicle CAN networks.
They mentioned it is possible for audio files to be spread though peer-to-peer networks.

Against those demonstrations, we propose a novel method that solves the fundamental security problems associated with the in-vehicle CAN network.
Our method works in the CAN physical layer and extracts characteristics of each ECU from measured analog signals.
Even if security measures suggested by previous researchers could be adapted to some new ECU design, the effectiveness of which is uncertain, our method has the distinct advantage of being compatible with the existing CAN system. In particular, we do not require existing ECUs to be replaced with new ones, and our method comports with the existing state of CANs. Our work can be used for such security purposes as intrusion detection (discovering ECU impersonation and network tampering), authentication (preventing unauthorized access to the physical network), forensic data collection (tying a physical device to a specific network incident), and quality assurance monitoring (determining whether a device will or is in the process of failing).

\section{Backgrounds}
\label{Backgrounds}
In this section, we describe device inconsistencies inherent in ECUs and how Controller Area Networks (CANs) function, in order to provide background necessary to understand our method.

\subsection{Inconsistency of Device Signals}
We have already mentioned that message authentication is actually impossible in the in-vehicle CAN network.
Hence, we do not seek to use message authentication, but rather the goal of our work is to identify ECUs by examining the distinctive analog and digital characteristics of these devices.
By differentiating ECUs based on the inimitable characteristics of signals emitted by individual ECUs, we can identify which ECUs are the original ones associated with the vehicle.
Signals with other characteristics would therefore be considered to be attacks from a malicious adversary emanating from other ECUs in the vehicle that have become hostage to an adversary or from alien ECUs planted by the adversary in the vehicle.
This method makes use of ECU hardware and manufacturing inconsistencies.
Figure \ref{An example of variations in signaling} shows an example of inherent variations in signaling behavior of CAN transceivers.
Although two transceivers might appear to be identical products from the same vendor and sending the exact same message, they generate different signals.
Those inconsistencies cause minute and unique variations in the signaling behavior of every digital device \cite{Wired_physical_layer}.
There are even minute variations in signals sent from the very same ECU in a particular vehicle.
In general, the variations are within a defined range and the unique characteristics of signals from individual ECUs remain constant over time; therefore, ECUs can be identified by their unique signals.

\subsection{CANs (Controller Area Networks)}
\label{CANs}
CAN is a communication protocol which was developed in the mid-1980s by Bosch GmbH \cite{CAN_Standard}.
CAN was first designed to provide a cost-effective communication bus for automotive applications, but is today widely used in various industries such as aerospace and railways, elevators, and medical devices \cite{di2008understanding}.
It is necessary to understand the physical layer of CAN, because our method deals with characteristics in the physical layer.
Since today's in-vehicle communications are designed based on the CAN 2.0 standard, we explain the CAN standards in terms of i) encoding / data transmission, ii) data frame formats, and iii) arbitration. In general, the in-vehicle CAN network topology is divided into two groups: the high-speed CAN and low-speed CAN (fault-tolerant CAN). The low-speed CAN consists of ECUs for easy functions such as the door open / lock function. Other more critical functions, such as engine or brake functions, are connected to the high-speed CAN. In this paper, we only consider the high-speed CAN, because this presents the most challenging environment for identifying signal characteristics \cite{Wired_physical_layer}.
Accordingly, once we show that our method is able to identify characteristics at a higher bit rate, it will follow that it will also be able to do so at a lower bit rate.
%%%%%%%%%%%%%%%%%%%%%%%%%%%%%%%%%%%%%%%%%%%%%%%%%%%%%%%%%%%%%%%%%%%%%%%%%%%%%%%%%%%%%%%%%%%%%%%%%%%%%%%%
\begin{figure}[h]
\centering
\includegraphics[width=3.5in]{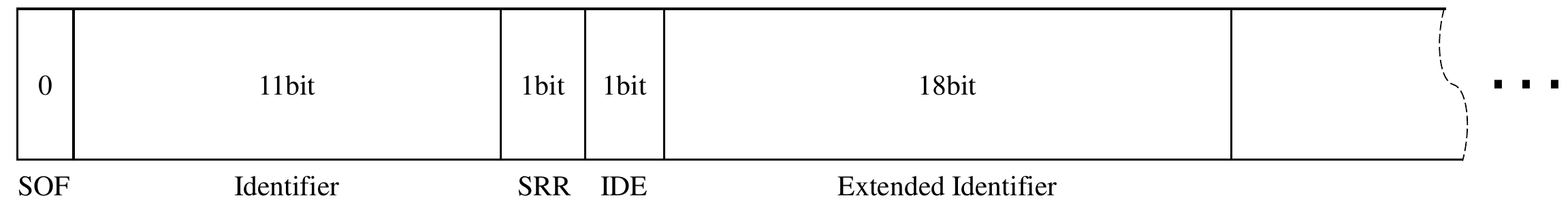}
\caption{CAN data frame format (Extended frame)}
\label{Fig: CAN data frame format}
\end{figure}
%%%%%%%%%%%%%%%%%%%%%%%%%%%%%%%%%%%%%%%%%%%%%%%%%%%%%%%%%%%%%%%%%%%%%%%%%%%%%%%%%%%%%%%%%%%%%%%%%%%%%%%%

\subsubsection{Encoding / Data Transmission}
\vspace{0.3cm}
The CAN signal is encoded using the Non Return to Zero (NRZ) bit encoding method, in which 1 is represented by one particular voltage and 0 is represented by some other significant voltage. A twisted pair of two wires in a shielded cable is used for Can communication. Both lines, CAN-H (High) and CAN-L (Low) are biased at 2.5 volts in the case of the recessive state (1). For the dominant state (0), CAN-H goes to around 3.5 volts and CAN-L to 1.5 volts, respectively. In terms of device inconsistencies, those degrees of voltage are all different for different devices. From those differences, we determine the signal characteristics that correspond to particular ECUs.

\subsubsection{Data Frame Format}
\vspace{0.3cm}
The CAN defines four types of frames: i) data frame, ii) remote frame, iii) error frame, and iv) overload frame.
Of these four types, we only describe the data frame, because the purpose of our method is to identify which ECU is transmitting the data frame.
As an interesting features of the CAN, the identifier field of the data frame refers to the identifier of the transmitter ECU not that of the receiver.
Accordingly, it is possible to determine which ECU is transmitting the data frame by checking the identifier.
Our goal is to detect an impersonator and identify the ECU from which that impersonator is transmitting, using its unique signal characteristics.
There are two types of data frames: one is the base frame format whose identifier length is 11 bits, and the other is the extended frame format, which has 29 identifier bits made up of the 11-bit identifier and an 18-bit identifier extension. We refer to the extended identifier as EXID. To use the extended frame format, the identifier extension bit (IDE) is transmitted in the recessive (1) state. Figure \ref{Fig: CAN data frame format}, shows an overview of the CAN data frame and we describe each field in the data frame in Table \ref{Fields of CAN data frame.}. Our method measures signals corresponding to the bit string of the extended identifier field. The reason why we do not use the signals corresponding to the identifier field is because of arbitration. We will describe arbitration below.

%%%%%%%%%%%%%%%%%%%%%%%%%%%%%%%%%%%%%%%%%%%%%%%%%%%%%%%%%%%%%%%%%%%%%%%%%%%%%%%%%%%%%%%%%%%%%%%%%%%%%%%%
\begin{table}[t]
%\small
\centering
\captionsetup{justification=centering,margin=0.5cm}
\caption{Fields of extended frame format}
\label{Fields of CAN data frame.}
\begin{tabular}{|p{1.8cm}|c|p{4cm}|}
\hline
Field & Length   &   Description \\
        &   (bits)  &   \\\hline
Start-of-frame  &   1   & Means the start of frame transmission \\ \hline
Identifier  &   11   & An identifier of transmitter, which also represents the message priority \\ \hline
Substitute remote request (SRR)  &  1    & Must be recessive state (1) \\ \hline
Identifier extension bit (IDE)  &   1   &  Indicating whether 11 bit identifier or 29 bit extended identifier is used. Dominant state (0) indicate 11 bit identifier while Recessive state (1) indicate 29 bit extended \\ \hline
Extended identifier (EXID)  &  18    &  If IDE is Recessive state (1), EXID field is available\\ \hline
Reserved bits    &   2   &  $~~~~~~~~~~~~~~~~~~$ -\\ \hline
Data length code (DLC)    &   4   &   Number of bytes of data (0-8 bytes)   \\ \hline
Data field  &   0-64    &   Data to be transmitted\\ \hline
CRC &   15  &  Cyclic redundancy check \\ \hline
CRC delimiter   &   1   &   Must be recessive state (1)\\ \hline
ACK slot    &   1   &  Transmitter sends recessive state (1) and any receiver can assert a dominant sate (0) \\ \hline
ACK delimiter    &   1   &  Must be recessive state (1) \\ \hline
End-of-frame    &   7   &  Must be recessive sate (1) \\ \hline
\end{tabular}
\end{table}
%%%%%%%%%%%%%%%%%%%%%%%%%%%%%%%%%%%%%%%%%%%%%%%%%%%%%%%%%%%%%%%%%%%%%%%%%%%%%%%%%%%%%%%%%%%%%%%%%%%%%%%%

%%%%%%%%%%%%%%%%%%%%%%%%%%%%%%%%%%%%%%%%%%%%%%%%%%%%%%%%%%%%%%%%%%%%%%%%%%%%%%%%%%%%%%%%%%%%%%%%%%%%%%%%

\subsubsection{Arbitration / Extended Identifier}
\begin{figure}[!t]
\centering
\includegraphics[width=3.5in]{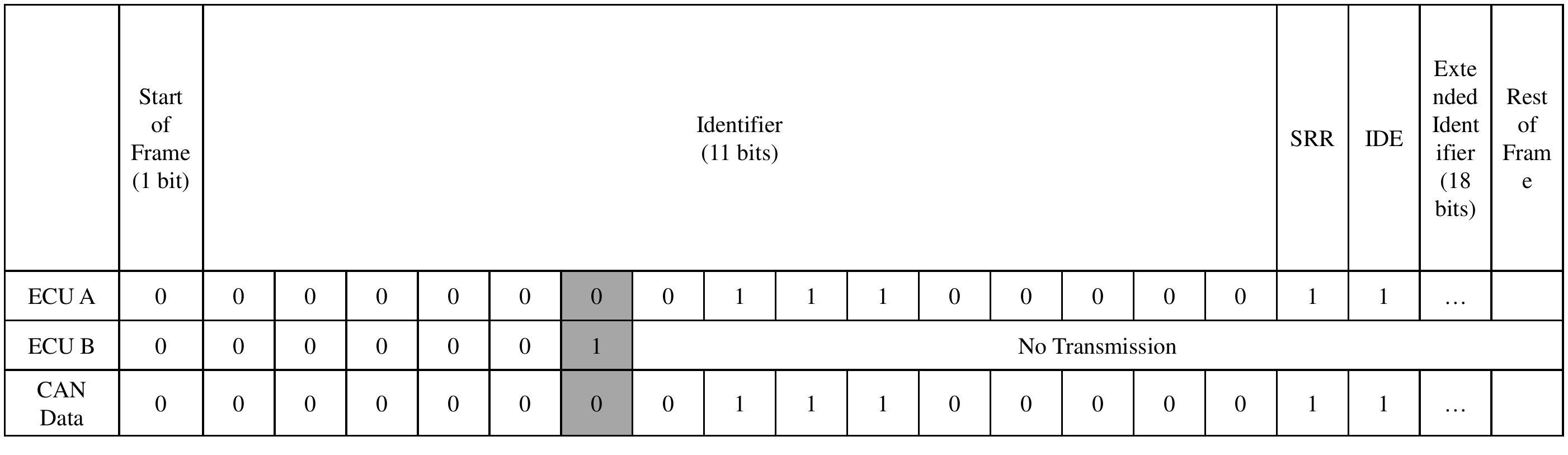}
\caption{An example of arbitration between two nodes}
\label{An example of arbitration between two nodes}
\end{figure}

\vspace{0.3cm}
There are times when more than two ECUs each transmit data at the same time.
To prioritize such collision signals, CAN supports a lossless bit-wise arbitration method.
If one ECU transmits a dominant bit (0) and another ECU transmits a recessive bit (1), then there is a collision and the dominant bit gets higher priority.
During transmission, an ECU continuously checks on the bus and compares the signal states, recessive or dominant.
Since the identifier field is the first field in the data frame, the identifier field is important in terms of the arbitration decision.
The priority of ECU signals depends on their identifiers. Figure \ref{An example of arbitration between two nodes} shows an example of how one of two ECU signals gets higher priority through arbitration.
The two ECUs have different identifiers for the first time in the 6th bit of the identifier, where ECU A is dominant (0) and B is recessive (1).
ECU A gets higher priority, and ECU B's signal is cut off and can try again later.
Before an arbitration decision, multiple signals in the identifier field are generated from multiple ECUs.
In our example in Figure \ref{An example of arbitration between two nodes}, the first five parts of the signals being sent from the two ECUs are the signal generated by both ECUs.
Accordingly, it is impossible to identify the unique characteristics of each single ECU from the signals in the identifier field.
The extended identifier field was originally created in the CAN to accommodate situations requiring identification of more connection devices than the identifier field could accommodate.
The identifier field in the CAN is able to assign identifiers to many more connection devices than the 50-70 ECUs found in vehicles, since the identifier field is capable of identifying thousands of connection devices at a time.
The increased capacity of the extended identifier field was needed for other industry applications. However, in the case of in-vehicle ECUs, the deficiency of the identifier field is not in the number of connection devices it can potentially identify, but rather is in its incapacity to differentiate the accurate ECU origin of signals under collision conditions.
Hence, our activation and use of the extended identifier field is not to increase the number of potential identifiers, as it was originally created to do, but rather to overcome the problem of signal identification created by signal collisions.
Since there are only 50-70 ECUs in a vehicle and the identifier field will have been sufficient to handle all arbitration decisions, we therefore programmed the extended identifier field to be included into the data frame along with the content of the message.

%%%%%%%%%%%%%%%%%%%%%%%%%%%%%%%%%%%%%%%%%%%%%%%%%%%%%%%%%%%%%%%%%%%%%%%%%%%%%%%%%%%%%%%%%%%%%%%%%%%%%%%%
%%%%%%%%%%%%%%%%%%%%%%%%%%%%%%%%%%%%%%%%%%%%%%%%%%%%%%%%%%%%%%%%%%%%%%%%%%%%%%%%%%%%%%%%%%%%%%%%%%%%%%%%
\section{System Model}
\label{System}
%%%%%%%%%%%%%%%%%%%%%%%%%%%%%%%%%%%%%%%%%%%%%%%%%%%%%%%%%%%%%%%%%%%%%%%%%%%%%%%%%%%%%%%%%%%%%%%%%%%%%%%%
\begin{figure}[!t]
\includegraphics[width=3.5in]{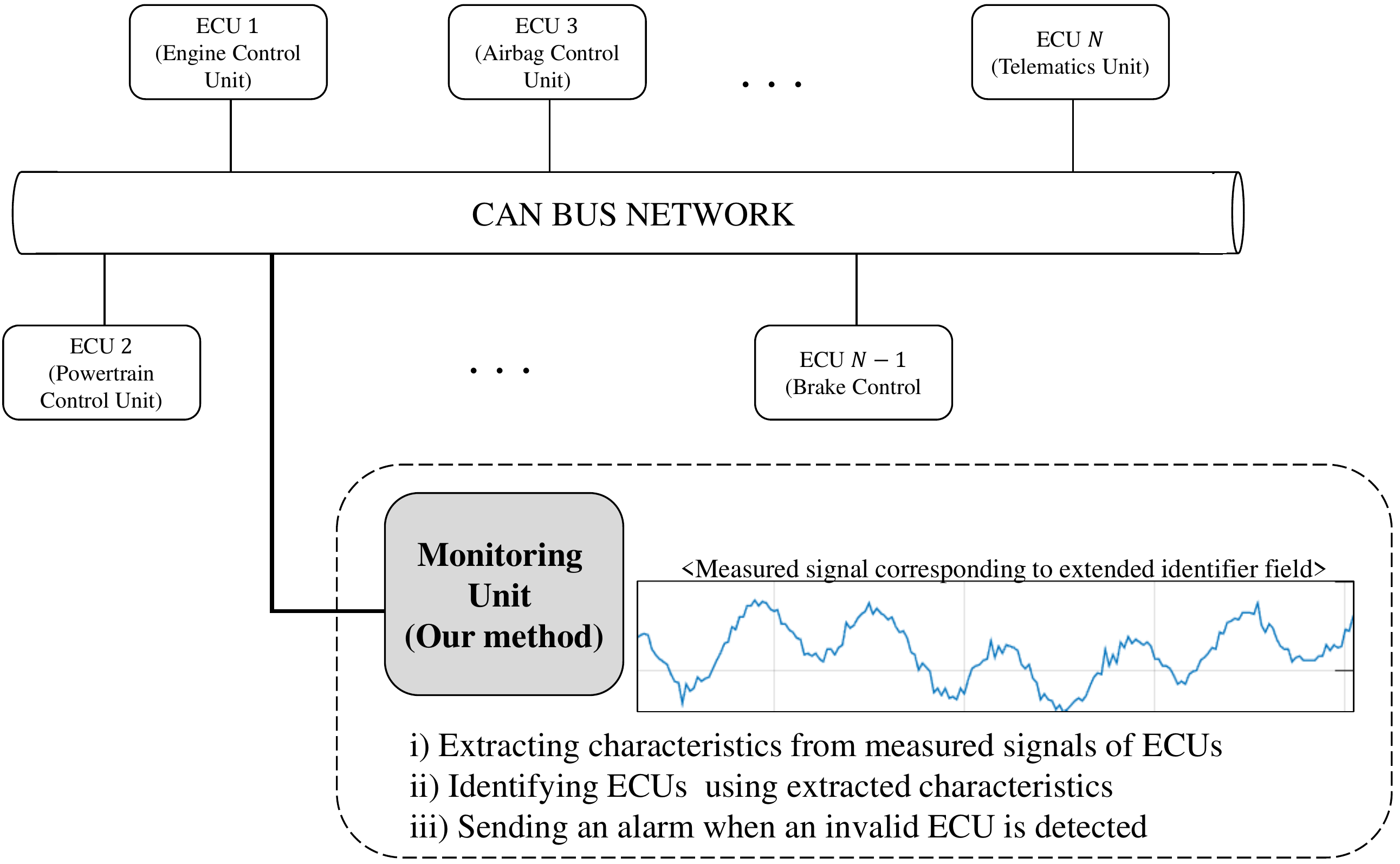}
\caption{System Model}
\label{System model}
\end{figure}
%%%%%%%%%%%%%%%%%%%%%%%%%%%%%%%%%%%%%%%%%%%%%%%%%%%%%%%%%%%%%%%%%%%%%%%%%%%%%%%%%%%%%%%%%%%%%%%%%%%%%%%%
In this section, we present a system model for our method. Figure 4 shows how our method works in order to identify ECUs in the in-vehicle CAN network.
As shown in Figure \ref{System model}, our system model adds a monitoring unit to the CAN.
The monitoring unit is programmed to analyze sent messages in several ways.
First, it extracts the known ECU identifier from the identifier field and determines whether the identifier is a known identifier or not.
Second, the monitoring unit analyzes the features of the signals in a sent message with the enabled extended identifier field.
Basically, the monitoring system utilizes a classification algorithm to which the same message is sent multiple times so that the algorithm can learn them to create a fingerprint template of the unique characteristic from each ECU. Hence, the unique signal characteristics of each ECU are contained in fingerprints as determined by our fingerprinting method. For example, the monitoring unit will know that signals to turn on and off the engine are sent by ECU 1, while signals to control the brakes are sent by ECU 2. The monitoring unit is programmed to know which identifiers should match which fingerprints: it has both ECU identifiers and signal fingerprints in correct pairings.

The monitoring unit receives signals sent by all ECUs, and since it knows which signals should be coming from which ECU, it will know when signals are sent by the wrong ECU. In the case of an adversary attack, the adversary can use the identifier of the ECU that the adversary knows should be sending the signal (telling the engine to shut off, for example). When that happens, the monitoring unit will know that the signal did not originate from the ECU that it is paired with, the one that is supposed to be generating that signal. This will identify the presence of an adversary and the mismatched fingerprint will identify the ECU that was actually used by the adversary to send the signal, if it is an existing ECU that has been taken hostage by the adversary, or it will indicate that the signal originated from some other completely unknown device. In other words, when a particular signal is not being generated from the properly assigned ECU identifier, the monitoring system knows it and it then sends an alarm to the vehicle owner.

Because CAN bus ECUs are all connected to the same shared bus line, we can apply our method to current vehicle systems by adding the monitoring unit to the in-vehicle CAN network and adjusting ECU programming to include the extended identifier field in ECU signals. To clarify our method, we will describe some possible adversary models and our underlying assumptions about the CAN.

\subsection{Adversary Models}
\label{Adversary_model}
The goal of adversaries is to transmit malicious commands to the in-vehicle CAN network so that the adversary can take control of the vehicle.
Because the CAN does not support message authentication, an adversary could simply attack by connecting with the in-vehicle CAN network and launching a replay attack.
We consider two types of adversary threats based on how the adversary would be able to access the in-vehicle CAN network.

\textbf{Adversary Model Type I.} The first adversary method is one where the adversary gains access through an additional external device that is physically planted by the adversary. The new device could be inserted into the On-Board Diagnostics (OBD)-II port, which is an interface between in-vehicle networks, including the in-vehicle CAN network. OBD systems give the vehicle owner or repair technician access to the status of various vehicle subsystems. Various tools are available that plug into the OBD connector to access in-vehicle sub-networks, including in the CAN network. Since the OBD-II port is located under the dashboard, an adversary could gain access to the in-vehicle CAN network by attaching an additional device to the OBD-II port.  The adversary could also gain access through an external device (e.g., laptop or smartphone). A mobile device-based or PC-based scan tool is one that an adversary can easily buy on the Internet \cite{ELM327,OBD2USB}.
Using such additional devices, an adversary is able to transmit malicious commands.

\textbf{Adversary Model Type II.} The second adversary model is for the adversary to compromise an existing ECU in the in-vehicle CAN network.
In this attack, an adversary basically takes an existing ECU hostage to imitate existing signal features and send messages with the correct signal but from a different ECU that they have been able to take hostage.
To root out attackers who are operating from a hostage ECU, we needed to identify signal fingerprints that are not paired with the correct ECU identifier.
One example would be compromise of a telematics ECU. Telematics ECUs are being installed in more and more vehicles to enable additional functions (e.g., remote engine start through a smartphone).
Due to the fact that telematics ECUs provide several access points of access through external networks such as Bluetooth, Wi-Fi, and 3G communication, telematics ECUs are vulnerable to attack by the adversary.
In fact, Chekoway et al. performed such an attack on a telematics ECU using these known vulnerabilities \cite{CarsharkII}.
After compromising a telematics ECU, the adversary is then able to transmit messages to intentionally cause malfunctions.

We separate adversary models according to the way an adversary is able to connect to the in-vehicle CAN network. We consider Adversary Model Type I to constitute a lesser threat than Adversary Model Type II. For one thing, it is difficult for a Type I adversary to succeed since it requires additional devices to be plugged into the OBD-II port that can easily be seen by the driver and removed. Since Type I is so much more difficult to localize and eradicate, we designed our method to be effective with a Type II adversary.  Our method can be used to address both types of adversaries

\subsection{Our Assumptions}
\label{Our_assumptions}
Our method uses the extended frame format instead of the base frame format, since the base frame format with only its identifier field does not properly identify signal characteristics in collision situations, described in Section 3.
The extended frame format contains both the 11-bit identifier field and also an 18-bit extended identifier field. We assume that all arbitration decisions are made within the identifier field since we know that it has the capacity to accommodate all signals sent by the 50-70 some vehicle ECUs. Based on the assumption that all collision signals will have been subject to arbitration decisions in the identifier field, we further assume that signals in the extended identifier field are not affected by arbitration decisions so that the unique characteristics of signals from different ECUs will be distinguishable. In other words, we assume that the extended identifier field will allow measurement of physical signals of all sent messages without interference from the effects of arbitration that render the identifier field by itself to be unreliable in accurately distinguishing those characteristics.
The second assumption is that it is impossible to imitate an ECU's unique signal characteristics. Due to the effects of hardware inconsistencies, there are subtle differences between devices in terms of signal variation. Thus, we assume that those variations are unique characteristics of devices and inimitable. Although an adversary could learn ECU identifier, the adversary could not hide the fact that a message is being sent from an alien or from a hostage ECU.
Finally, we assume that a single malicious command will not be effective to cause vehicle malfunction. For example, if an adversary wants to incapacitate the brake system, he must transmit malicious messages over and over until the intended malfunction occurs. This means that a single malicious message is assumed to be ineffective. We therefore assume that an adversary will transmit malicious messages repeatedly in order to perpetrate an attack on a vehicle.

\section{Our Method}
We describe the method we developed to identify ECUs in an in-vehicle CAN network.
We utilized the characteristics of CAN signals that are generated from the CAN transceiver from each ECU.
Our main idea was to extract suitable statistical features of signals, which include unique characteristics, and then classified these unique signals into Adversary Model Type I or Adversary Model Type II.  A detailed description of our method follows.

\subsection{Assigning Bit String / Data Collection}
We measured the physical signal transmitted by an ECU as identified in the extended identifier field in the CAN data frame.
First of all, we defined the extended identifier as an 18-bit string, so that every signal in the CAN data frame would have the same length bit string in the extended identifier field.
Moreover, the most significant bit (MSB) of the bit string was 0 to prevent bit stuffing at a different location.
In the CAN network, six consecutive bits of the same type are considered to be an error, so that bit stuffing automatically occurs in the CAN network when one bit of the opposite polarity is inserted after five consecutive bits of the same polarity.
The other 17 bits could be any value (i.e., 0 or 1) as follows:
\begin{equation} EXID = b_1b_2 b_3b_4b_5b_6 b_7b_8b_9b_{10} b_{11}b_{12}b_{13}b_{14} b_{15}b_{16}b_{17}b_{18}, \end{equation}
where $b_1=0$ and $b_i\in\{0,1\}, $~~~$ 2\leq i\leq18$.
We measured a signal corresponding to the extended identifier field.
The sampled signal was denoted by $S(k)$, a set of sample values.

\subsection{Fingerprint Generation}
We took the method proposed by S. Dey et al. \cite{Accelprint} in order to extract suitable statistical features from the sampled signal $S(k)$.
For the sampled signal $S(k)$, we extracted 40 scalar features in both time and frequency domains using LibXtract, a popular feature extraction library \cite{LibXtract}.
Among the 40 possible features, we selected the only relevant features by ranking the features using the FEAST toolbox and utilize the joint mutual information criterion for ranking \cite{FEAST}.
From the results, we selected the top 8 time domain features and 9 frequency domain features.
Table \ref{A list of time domain features} and Table \ref{A list of frequency domain features} show the selected 17 features. As a result, our method extracted those 17 features from a sampled signal, then the following steps were performed to identify individual ECUs, each with its own unique identifying characteristics.

%%%%%%%%%%%%%%%%%%%%%%%%%%%%%%%%%%%%%%%%%%%%%%%%%%%%%%%%%%%%%%%%%%%%%%%%%%%%%%%%%%%%%%%%%%%%%%%%%%%%%%%%
\begin{table} [h]
%\small
\centering
\captionsetup{justification=centering,margin=0.5cm}
\caption{A list of time domain features. Vector $x$ is the time domain representation of the data. $N$ is the number elements in $x$
} \label{A list of time domain features}
\begin{tabular}{|c|c|}
\hline
Feature Name & Description \\ \hline
Mean &  $\hat{x}=\frac{1}{N}\sum_{i=1}^Nx(i)$  \\ \hline
Std-Dev  &  $\sigma=\sqrt{\frac{1}{N-1}\sum_{i-1}^N(x(i)-\overline{x})}$  \\ \hline
Average Deviation  &  $D_{\overline{x}=\frac{1}{N}\sum_{i=1}^N|x(i)-\overline{x}|}$  \\ \hline
Skewness  & $\gamma=\frac{1}{N}\sum_{i-1}^{N}(\frac{x(i)-\overline{x}}{\sigma})^3$   \\ \hline
Kurtosis  &  $\beta=\frac{1}{N}\sum_{i=1}^N(\frac{x(i)-\overline{x}}{\sigma})^4-3$  \\ \hline
RMS Amplitude  & $A=\sqrt{\frac{1}{N}\sum_{i=1}^N(x(i))^2}$   \\ \hline
Lowest Value  & $L=(Min(x(i))|_{i=1..N})$   \\ \hline
Highest Value  & $H=(Max(x(i))|_{i=1..N})$   \\ \hline
\end{tabular}
\end{table}
%%%%%%%%%%%%%%%%%%%%%%%%%%%%%%%%%%%%%%%%%%%%%%%%%%%%%%%%%%%%%%%%%%%%%%%%%%%%%%%%%%%%%%%%%%%%%%%%%%%%%%%%
%%%%%%%%%%%%%%%%%%%%%%%%%%%%%%%%%%%%%%%%%%%%%%%%%%%%%%%%%%%%%%%%%%%%%%%%%%%%%%%%%%%%%%%%%%%%%%%%%%%%%%%%
\begin{table} [h]
\centering
%\small
\captionsetup{justification=centering,margin=0.5cm}
\caption{A list of frequency domain features. Vector $y$ is the frequency domain representation of the data. Vectors $y_m$ and $y_f$ hold the magnitude coefficients and bin frequencies respectively. $N$ is the number of elements in $y_m$ and $y_f$.} \label{A list of frequency domain features}
\begin{tabular}{|c|c|}
\hline
Feature Name & Description \\ \hline
Spec. Std-Dev   &  $\sigma_s=$\\
                &  $\sqrt{(\sum_{i=1}^N(y_f(i))^2*y_m(i)))/(\sum_{i=1}^Ny_m(i))}$  \\ \hline
Spec. Skewness  &  $\gamma_s=(\sum_{i=1}^Ny_f(i)y_m(i))/\sigma_s^3$  \\ \hline
Spec. Kurtosis  &  $\beta_s=(\sum_{i=1}^N(y_m(i)-C_s)^4*y_m(i))/\sigma_s^4-3$  \\ \hline
Spec. Centroid  &  $C_s=(\sum_{i=1}^Ny_f(i)y_m(i))/(\sum_{i=1}^Ny_m(i))$  \\ \hline
Irregularity-K  &  $IK_s=$\\
                &  $\sum_{i=2}^{N-1}|y_m(i)-\frac{y_m(i-1)+y_m(i)+y_m(i+1)}{3}|$  \\ \hline
Irregularity-J  &  $IJ_s=\frac{\sum_{i=1}^{N-1}(y_m(i)-y_m(i+1))^2}{\sum_{i=1}^{N-1}(y_m(i))^2}$  \\ \hline
Roll off  &  $R_s=\frac{SampleRate}{N}*n|_{\sum_{i=1}^n y_m<Threshold}  $\\ \hline
Flatness  &  $F_s = (\prod_{i=1}^Ny_m(i))^{\frac{1}{N}}/((\sum_{i=1}^Ny_m(i))/N)$  \\ \hline
Smoothness  &  $S_s=\sum_{i=2}^{N-1}|20.log(y_m(i))-$\\
            &   $\frac{(20.log(y_m(i-1))+20.log(y_m(i))+20.log(y_m(i+1)))}{3}|$   \\ \hline

\end{tabular}
\end{table}
%%%%%%%%%%%%%%%%%%%%%%%%%%%%%%%%%%%%%%%%%%%%%%%%%%%%%%%%%%%%%%%%%%%%%%%%%%%%%%%%%%%%%%%%%%%%%%%%%%%%%%%%

A set of selected features of $S(k)$ is denoted as $F(S)$. As a result, we extracted each set of features $F(S)$ corresponding to a measured signal $S(k)$. Accordingly, $F(S)$ represents the fingerprint of an ECU.

\subsection{Fingerprint template generation}
ECU signals are as distinctive as fingerprints.
We needed some way to leverage the set of distinctive signal features to permit identification of the ECU generating that particular set of features.
To achieve this goal, we used a classification algorithm which is designed for the problem of identifying to which of a set of categories a new observation belongs, on the basis of a training set of data containing observations whose category membership is known. In this paper, the concept of using a classification algorithm is to train a classifier with a lot of CAN signals that valid ECUs sends and then, the signal that a malicious ECU sends can be identified by the trained classifier.
We note that classification algorithms are already broadly utilized as a useful tool for security methods. For example, IDSs (Intrusion Detection Systems) can be designed with a classification algorithm to detect an anomaly behavior in a network \cite{garcia2009anomaly}. Accordingly, we can say that our method with a classification algorithm is also acceptable.
To generate ECU fingerprint templates, the first step was to create a classifier by using observations from ECUs as inputs.
The observations were the sets of 17 features for each ECU, that is, the $F(S)$ of each ECU.
The classifier was then trained in a supervised machine learning environment so that it could later be used to compare new observations to the previously-observed sets.
Our goal was to train the classifier to recognize each signal's features, its fingerprint, to use as a reference to later match with the identifier for the ECU that should be generating that fingerprint. Later, if the signal fingerprint does not match that ECU identifier, we can conclude that the fingerprint is from an alien source and can send an alarm to the vehicle owner (Type I adversary model).
Even if a fingerprint matches a valid ECU identifier, it should be checked if the identified ECU is allowed to send a command which is one of the CAN messages.
When it is detected for a ECU to send a command that the ECU is not allowed to send, this means that the ECU is compromised by an adversary (Type II adversary model).

The classifier is trained with n sets of features for each ECU.
In our case, we designed twelve CAN development boards to simulate standard ECUs.
In order to train the classifier to recognize all twelve ECUs, we first created fingerprints by inputting 900 observations in the form of 900 signals sent from each ECU.
Although the same message was being sent from the same ECU, each signal observation would have slight variations of the 17 features of that ECU's signals.
Through this supervised learning, we trained the classifier to average the slight variations in each ECU's signals into twelve fingerprint templates. This procedure would typically be performed to produce fingerprint templates at the time of manufacture of a vehicle.
And, it would need to be performed again whenever a new ECU is added to or an ECU is removed from the in-vehicle network; however, we note that adding, removing, or replacing an ECU after initial manufacture is an infrequent occurrence in an in-vehicle network, so that this learning phase is not frequently performed after leaving the factory.

\subsection{Fingerprint Matching}
After completing the above learning phase, \emph{Fingerprint template generation}, which produced fingerprint templates for existing ECUs in the vehicle, we then proceeded to the testing phase and matched new signals with the previously learned fingerprints.
Two types of fingerprint matching were performed. First, we tested for signs of Type I adversaries. Since Type I adversaries utilize additional devices that are not identified in the known database of ECU identifiers and therefore send signals that are not in the monitor's database of known signal fingerprints, the monitoring unit needs to determine whether there is a sufficient match with any of the known fingerprints or not. If the monitoring unit can not match new signals with any known signals fingerprint templates, the inquiry is over. It means there is an adversary with a completely alien signal from a completely alien device. The alien signal identification belongs to novelty detection which is the identification of new or unknown data that a machine learning system has not been trained with \cite{muller2001introduction}. We presented a simple threshold-based approach for alien detection problem, which is a convenient extension of our classification-based basic model.

Comparing a new signal with a given feature set F(S) to all known fingerprint templates in the classifier yields scores showing the new feature set F(s) belongs to the known one of the fingerprint templates. Thus, if all scores are below the threshold, we classify the feature set F(S) to the class of unknown sets (i.e., an invalid ECU or device is the source of the signal). In short, we set a threshold, explained in Subsection \ref{Identifying_unknown_ECUs}, below which the monitoring unit determines that a newly identified fingerprint is an unknown adversary. An alarm will be sent.

By contrast, for Type II attacks we can directly apply a probabilistic model to determine which fingerprint in the known database a new signal most closely matches. Once we have determined that there are no alien ECUs, we can then determine if signals are being sent by an ECU other than the one properly paired with its signal fingerprint. To accomplish this, we produced fingerprints of new signals and matched them against known identifiers in Adversary Model Type II to ensure that they were consistent with known fingerprints. In this testing phase, the monitoring system searched the database of known fingerprints and matched the new fingerprint based on the probabilistic model; that is, it determined the known fingerprint with the highest probability of being the same as the new fingerprint above all others in the known database. It then determined whether that fingerprint originated from the correct ECU identifier programmed into the monitoring unit for that fingerprint. A mismatch resulted in an alarm to the owner.

We note that the essential difference between the two types of classifications is whether a threshold is used to determine that a new fingerprint is not in the known database of fingerprints at all, leading to the conclusion that there is an external malicious device (Type I), or whether a probabilistic model is used to determine whether a known signal fingerprint has been sent from an alien ECU within the CAN, leading to the conclusion that an existing ECU has been taken hostage to send malicious messages from another ECU in the CAN (Type II).

\section{Experimental Results}
We evaluated our method by performing a series of experiments. We assessed both our success rate in detecting alien signals (Adversary Type I) and in identifying correctly or incorrectly matched pairs of signal features and ECUs (Adversary Type II). Our experiments yielded success rates in both of over 90$\%$.
This success rate should be trustworthy to identify adversaries since, as we mentioned in section 4.2 above, successful adversary attacks are expected to occur in the in-vehicle CAN network only after transmitting multiple malicious signals. Accordingly, the probability that an adversary will succeed in an attack gets much lower because the adversary must be classified as within the class of known ECUs many times in order for the system to be duped into accepting an adversary message or source as a legitimate one. We use typical metrics common to machine learning and data mining are used to evaluate our method \cite{Glossary_of_Terms}.

\subsection{Experimental Setup}

\begin{figure}[!t] \centering
\includegraphics[width=3in]{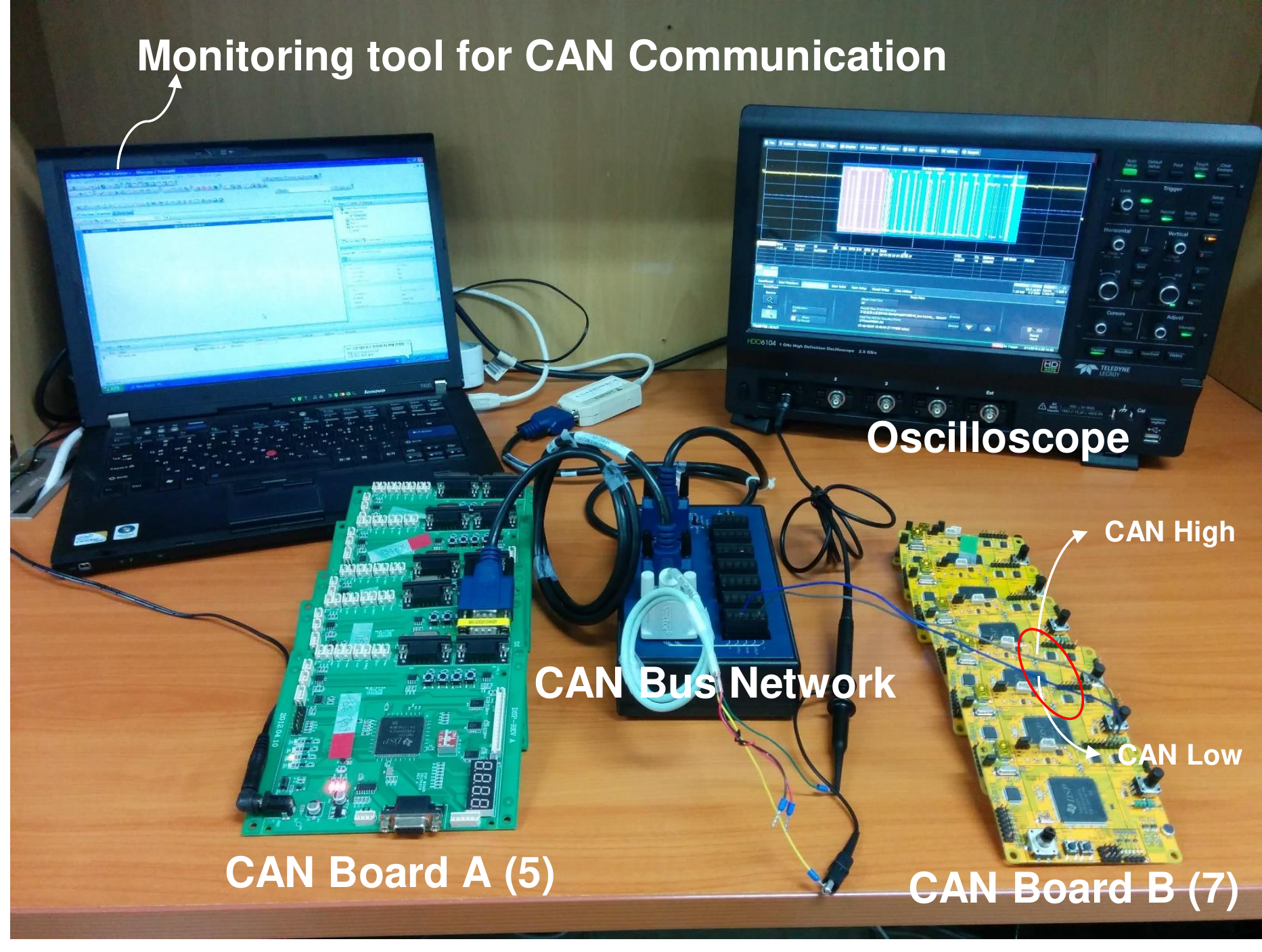}
\caption{Experimental setup}
\label{Fig: Experimental setup}
\end{figure}

We tested our method with more than 900 input observations, which were signals with the same message, sent  from each of 12 CAN development boards (shown as CAN Board A and B in Table \ref{Components of experimental setup}).
The sample signals were at the 500K bit rate typically used in the in-vehicle high-speed CAN.
In fact, there is another in-vehicle CAN network topology, a low-speed (fault tolerant) CAN whose bit rate is only 100K or 125K bits per second.
In this paper, we used only the high-speed CAN and its bit rate.
It is more difficult to identify devices using signal characteristics sampled at a high bit rate than at a low bit rate, so we assumed that if our method worked successfully at a high-speed CAN bit rate, then it would also function well at the low-speed CAN (fault tolerant) bit rate \cite{Wired_physical_layer}.
In addition, we used the default applications for classification provided by MatLab R2016a in order to use its classification algorithms \cite{MATLAB}.
The classification algorithms we used were Support Vector Machine (SVM), Neural Network (NN), and Bagged Decision Tree (BDT) with default options MatLab provides.
We performed the 10-fold cross validation test as a model validation technique for assessing how the results of a statistical analysis will generalize to an independent data set. In 10-fold cross validation, a given data set was partitioned into ten folds. A classifier was trained with nine folds (training data) and the classification model was tested with the remaining one fold (testing data). This process was repeated ten times so that each fold could be used exactly once as the validation data. We used the averaged classification accuracy as 10-fold cross validation accuracy (i.e., every result which we show in this section is the result of 10-fold cross validation test).
In Subsection \ref{Adversary_model} above, we mentioned two adversary models and we proceeded to evaluate our method on Adversary Model Type I and the other in a Type II framework.
Fig. \ref{Fig: Experimental setup} and Table \ref{Components of experimental setup} shows the purpose of using each component and its specification.
We note that even though the total number of ECUs is about 70 in case of a luxury sedan, in-vehicle networks are physically divided into several sub-networks.
Volvo XC90, for example, has 7-12 ECUs for an in-vehicle subnetwork \cite{Volvo_XC90}.

%%%%%%%%%%%%%%%%%%%%%%%%%%%%%%%%%%%%%%%%%%%%%%%%%%%%%%%%%%%%%%%%%%%%%%%%%%%%%%%%%%%%%%%%%%%%%%%%%%%%%%%%
\begin{table} [!h]
\centering
%\small
\captionsetup{justification=centering,margin=0.5cm}
\caption{Components of experimental setup}
\label{Components of experimental setup}
\begin{tabular}{|p{1.6cm}|p{3.4cm}|p{1.8cm}|}
\hline
Components & $~~~~~~~~$Specification & $~$Explanation \\ \hline
& HDO6104 (Lecroy) & For CAN   \\
& Bandwidth: 1GHz & data frame, it\\
Oscilloscope & Sampling rate: 2.5GS/s &  measures \\
& Vertical resolution: 12-bit &  voltage signal\\ \hline
CAN bus network  & CAN breakout box (National Instrument) & It provides CAN bus network \\ \hline
Monitoring tool for CAN Communication  & Off-the-shelf laptop $~~~$ CAN Communication monitoring tool: PCAN-Explorer & It monitors CAN packets on the CAN bus network \\ \hline
 & Order-made board  &  \\
  CAN development  board    & Microcontroller: TMS320F28335 (TEXAS INSTRUMENTS)  &  It transmits CAN data frame\\
   (Type A) & Transceiver: MCP2551 (MICROCHIP)  &  \\ \hline
     & TR28335 Training Kit (SyncWorks)  &  \\
 CAN development board       & Microcontroller: TMS320F28335 (TEXAS INSTRUMENTS)  & It transmits CAN data frame \\
(Type B)& Transceiver: SN65HVD235 (TEXAS INSTRUMENTS)  &  \\ \hline
\end{tabular}
\end{table}
%%%%%%%%%%%%%%%%%%%%%%%%%%%%%%%%%%%%%%%%%%%%%%%%%%%%%%%%%%%%%%%%%%%%%%%%%%%%%%%%%%%%%%%%%%%%%%%%%%%%%%%%

%%%%%%%%%%%%%%%%%%%%%%%%%%%%%%%%%%%%%%%%%%%%%%%%%%%%%%%%%%%%%%%%%%%%%%%%%%%%%%%%%%%%%%%%%%%%%%%%%%%%%%%%
\begin{table*}[!t]
  \centering
  \tiny
    \caption{Confusion matrix for identifying ECUs when SVM, NN, and BDT are used, respectively}
  \label{Table: Success rates of identification when SVM, NN, and BDT are used, respectively}
  \begin{tabular}{|c|p{0.7cm}|p{0.7cm}|p{0.7cm}|p{0.7cm}|p{0.7cm}|p{0.7cm}|p{0.7cm}|p{0.7cm}|p{0.7cm}|p{0.7cm}|p{0.7cm}|p{0.7cm}|}
\hline
      & ECU1                           & ECU2                           & ECU3                           & ECU4                           & ECU5                           & ECU6                           & ECU7                           & ECU8                           & ECU9                         & ECU10                      & ECU11                          & ECU12                      \\ \hline
ECU1&\textbf{98.22/ 98.33/ 97.78}&0/ 0/ 0.11&0/ 0/ 0&0.67/ 0.33/ 0.22&1.11/ 1.22/ 1.89&0/ 0/ 0&0/ 0/ 0&0/ 0.11/ 0&0/ 0/ 0&0/ 0/ 0&0/ 0/ 0&0/ 0/ 0\\\hline
ECU2&0/ 0/ 0&\textbf{95.89/ 95.11/ 95.89}&0/ 0/ 0&4.11/ 4.89/ 4.11&0/ 0/ 0&0/ 0/ 0&0/ 0/ 0&0/ 0/ 0&0/ 0/ 0&0/ 0/ 0&0/ 0/ 0&0/ 0/ 0\\\hline
ECU3&0/ 0/ 0&0/ 0/ 0&\textbf{99.33/ 98.78/ 98.67}&0.67/ 1.00/ 1.33&0/ 0.22/ 0&0/ 0/ 0&0/ 0/ 0&0/ 0/ 0&0/ 0/ 0&0/ 0/ 0&0/ 0/ 0&0/ 0/ 0\\\hline
ECU4&0.78/ 0.44/ 0.44&6.78/ 4.78/ 6.33&0.56/ 1.00/ 0.89&\textbf{91.67/ 93.78/ 91.67}&0.22/ 0/ 0.67&0/ 0/ 0&0/ 0/ 0&0/ 0/ 0&0/ 0/ 0&0/ 0/ 0&0/ 0/ 0&0/ 0/ 0\\\hline
ECU5&1.44/ 1.33/ 2.11&0/ 0/ 0&0.11/ 0.11/ 0.11&0/ 0/ 0.78&\textbf{98.33/ 98.44/ 97.00}&0/ 0/ 0&0/ 0/ 0&0/ 0.11/ 0&0.11/ 0/ 0&0/ 0/ 0&0/ 0/ 0&0/ 0/ 0\\\hline
ECU6&0/ 0/ 0&0/ 0/ 0&0/ 0/ `0&0/ 0/ 0&0/ 0/ 0&\textbf{86.33/ 85.00/ 86.33}&0/ 0/ 0&0.44/ 0.56/ 0.44&0/ 0/ 0&0/ 0/ 0&13.22/ 14.44/ 13.22&0/ 0/ 0\\\hline
ECU7&0/ 0/ 0&0/ 0/ 0&0/ 0/ 0&0/ 0/ 0&0/ 0/ 0&0/ 0/ 0&\textbf{99.78/ 99.78/ 99.56}&0/ 0/ 0&0/ 0/ 0&0.22/ 0.22/ 0.44&0/ 0/ 0&0/ 0/ 0\\\hline
ECU8&0/ 0/ 0&0/ 0/ 0&0/ 0/ 0&0/ 0/ 0&0/ 0/ 0&1.89/ 1.22/ 1.56&0/ 0/ 0&\textbf{97.00/ 97.56/ 95.22}&0/ 0/ 0&0/ 0/ 0&1.11/ 1.22/ 3.22&0/ 0/ 0\\\hline
ECU9&0/ 0/ 0&0/ 0/ 0&0/ 0/ 0&0/ 0/ 0&0/ 0/ 0&0/ 0/ 0&0/ 0/ 0&0/ 0/ 0&\textbf{100/ 99.89/ 99.44}&0/ 0.11/ 0.56&0/ 0/ 0&0/ 0/ 0\\\hline
ECU10&0/ 0/ 0&0/ 0/ 0&0/ 0/ 0&0/ 0/ 0&0/ 0/ 0&0/ 0/ 0&0/ 0/ 0.22&0/ 0/ 0&0/ 0/ 0.11&\textbf{100/ 100/ 99.00}&0/ 0/ 0&0/ 0/ 0.67\\\hline
ECU11&0/ 0/ 0&0/ 0/ 0&0/ 0/ 0&0/ 0/ 0&0/ 0.11/ 0&13.22/ 12.44/ 14.22&0/ 0/ 0&1.89/ 1.67/ 2.78&0/ 0/ 0&0/ 0/ 0&\textbf{84.89/ 85.78/ 83.00}&0/ 0/ 0\\\hline
ECU12&0/ 0/ 0&0/ 0/ 0&0/ 0/ 0&0/ 0/ 0&0/ 0/ 0&0/ 0/ 0&0/ 0/ 0&0/ 0/ 0&0/ 0/ 0&0/ 0/ 0.22&0/ 0/ 0&\textbf{100/ 100/ 99.78}\\\hline
\end{tabular}
\end{table*}
%%%%%%%%%%%%%%%%%%%%%%%%%%%%%%%%%%%%%%%%%%%%%%%%%%%%%%%%%%%%%%%%%%%%%%%%%%%%%%%%%%%%%%%%%%%%%%%%%%%%%%%%

%%%%%%%%%%%%%%%%%%%%%%%%%%%%%%%%%%%%%%%%%%%%%%%%%%%%%%%%%%%%%%%%%%%%%%%%%%%%%%%%%%%%%%%%%%%%%%%%%%%%%%%%
\begin{table*}[!t]
\centering
\captionsetup{justification=centering,margin=0.5cm}
\caption{Overall success rate corresponding to different bit string of extended identifer using three different classification algorithms: SVM, NN, and BDT} \label{Overall success rate corresponding to differen bit string of extended identifer using three different classification algorithms: SVM, NN, and BDT}
\begin{tabular}{|c|c|c|c|c|c|c|c|c|}
\hline
\multirow{3}{*}{EXID} & \multicolumn{2}{c|}{SVM}             & \multicolumn{3}{c|}{NN}              & \multicolumn{3}{c|}{BDT}            \\ \cline{2-9}
                      & \multicolumn{2}{c|}{Kernel function} & \multicolumn{3}{c|}{\# Hidden nodes} & \multicolumn{3}{c|}{\# Classifiers} \\ \cline{2-9}
                      & Linear             & RBF             & 10          & 50        & 100        & 10         & 50         & 100       \\ \hline
 $0000...0000_{(2)}$  & 94.94     &  91.75   &         95.78     &        96.41    &  \textbf{96.48}      & $95.40$ &   $95.79$     &  \textbf{95.92}       \\ \hline
 $0101...1010_{(2)}$  & \textbf{95.24}      &   $88.26$  &         $95.32$    &       $95.44$    &   $95.28$    &  $94.07$     &  $94.69$    &   $94.83$   \\ \hline
 $1111...1111_{(2)}$  & $76.50$     &  $76.59$   &         $79.51$    &       $82.71$     &   $82.13$    &   $82.25$    &  $84.20$    &   $84.71$     \\ \hline
\end{tabular}
\end{table*}
%%%%%%%%%%%%%%%%%%%%%%%%%%%%%%%%%%%%%%%%%%%%%%%%%%%%%%%%%%%%%%%%%%%%%%%%%%%%%%%%%%%%%%%%%%%%%%%%%%%%%%%%

\subsection{Basic Identification}
\label{Basic Identification}
Our method uses voltage signals corresponding to extended identifier field of the CAN data frame, which are generated by CAN transceivers of ECUs.
As a simple evaluation of our method, we define the 18-bit extended identifier as $010101010101010101_{(2)}$.
For the total 29-bit identifier, the most significant 11-bit identifier is used for standard identifiers and the least significant 18-bit identifier is used for extended identifiers of ECUs.
As we mentioned in Subsection \ref{CANs}, the identifier extension bit (IDE) is what triggers the inclusion of the extended identifier field in the signal.
When we define the IDE as 1, the extended frame format is enabled. Since ambiguous collision signals are assumed to occur only in the standard identifier field with arbitration decisions that then result in one signal never being sent, and since our monitoring unit only analyzes sent messages,  the signal corresponding to the extended identifier is an unambiguous signal generated by a single ECU. The extended identifier field, unlike the standard identifier field, does not involve multiple signals that collide and are the subject of arbitration and any confusion about proper identification of the bits in the signal are simply no longer a problem.

Using signals identified with the extended identifier field, we extract statistical features corresponding to each signal and then performed classification methods on the features. Table \ref{Table: Success rates of identification when SVM, NN, and BDT are used, respectively} is the confusion matrix which shows the success and misclassification rates using three typical classification methods. Each diagonal cell indicates the success rate for that signal or observation to be classified in the valid ECU, with three values representing the rates obtained from three different classification algorithms: SVM with linear kernel, NN with 100 hidden nodes, and BDT with 100 classifiers, in that order. It notes that we will show that theses parameters outperform the others in next subsection. The confusion matrix is a specific table layout that allow visualization of the performance of an algorithm, typically a supervised learning one. Each column of the matrix represents the instances in a predicted class, while each row represents the instances that the class was actually identified. The results show that our method has a higher success rate than that of Mervay et al. \cite{Cosic}.
We note that Mervay et al. obtained a misclassification rate of up to $99.9\%$ but our misclassification rate was up to $15\%$ and mostly lower than $5\%$. This result is only available for Type II adversary model. The result considering Type I adversary model will be mentioned in Subsection \ref{Identifying_unknown_ECUs}, Identifying unknown ECUs.  We conclude our method can accurately identify ECUs.

%%%%%%%%%%%%%%%%%%%%%%%%%%%%%%%%%%%%%%%%%%%%%%%%%%%%%%%%%%%%%%%%%%%%%%%%%%%%%%%%%%%%%%%%%%%%%%%%%%%%%%%%
\begin{figure*}[t]
\centering
\includegraphics[width=7in]{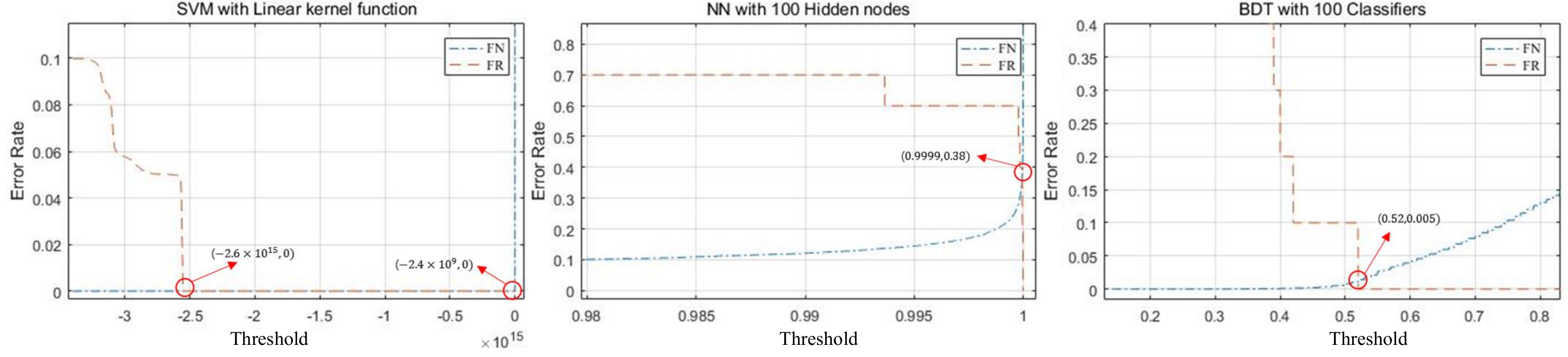}
\caption{Both FN and FP rate with varying threshold \\}
\label{Top - Overall Success Rate, Bottom - Alien Detection Rate}
\end{figure*}
%%%%%%%%%%%%%%%%%%%%%%%%%%%%%%%%%%%%%%%%%%%%%%%%%%%%%%%%%%%%%%%%%%%%%%%%%%%%%%%%%%%%%%%%%%%%%%%%%%%%%%%%

\subsection{Identifying known ECUs}
In the above subsection, we evaluated the success rates for classification when the 18-bit extended identifier is defined as $000000000000000000_{(2)}$.
In this subsection, we calculate the overall success rates while changing the extended identifier to other bit strings.
We want to know whether bit strings affect our success rate or not.
Thus, we additionally selected typical two bit strings because the number of times the state switches from  a recessive state to a dominant state, or the other way around, may have a significant effect on the success rate. The reason why the most significant bit we selected is 0 because of bit stuffing. Due to the fact that the IDE is 1 bit, we can avoid the case that bit stuffing occurs at different bit locations when we define the most significant bit of the extended identifier as 0 bit. Table \ref{Overall success rate corresponding to differen bit string of extended identifer using three different classification algorithms: SVM, NN, and BDT} shows the overall identification rates when the extended identifier is defined with a different bit string.
We used typical parameters which are used in three classification algorithm. We can see that linear kernel function, 100 hidden nodes, and 100 classifiers achieve the best performance of SVM, NN, BDT, respectively. In next subsection, we evaluate our method with only these parameters.

In addition, this shows that the success rate varies according to the bit string given to the extended identifier field.
Our method reaches the highest rate when the extended identifier was $000000000000000000_{(2)}$ and NN with 100 hidden layers was performed.
Even though we did not evaluate more bit string identifiers for the field, the success rates for both $000000000000000000_{(2)}$ and $010101010101010101_{(2)}$ are acceptable.
From these results, we conclude that we get better result when there are enough dominant states (i.e., 0 bit).

\subsection{Identifying unknown ECUs}
\label{Identifying_unknown_ECUs}
In the above subsection, we evaluated our method only considering Type II adversary model. To address a Type I adversary where an attacker attaches a new device to OBD-II port to connect to the in-vehicle CAN network, our method must classify the new device as belonging to the unknown class.
As we mentioned in Section V, identifying unknown ECUs is related to novelty detection which is the identification of new or unknown data that a machine learning system has not been trained with \cite{muller2001introduction}. We utilized a threshold-based approach which is a convenient and effective extension of our classification-based basic model. By using this approach, we can identify unknown ECUs while preserving our basic classification model.
Accordingly, we define a threshold and perform the same classification methods as in the above basic experiment. In the threshold model, we consider that signals whose scores corresponding to known classes are all below the threshold are from the unknown class. The overall success rates of the threshold-based identification are less or equal to ones of the basic identification without a threshold, because the case that classification scores is less than a threshold is classified as an unknown class, even if the result is correct.
FN (False Negative) case refers to the case that an valid ECU is classified as an alien (i.e., an unknown ECU) and FP (False Positive) case refers to the case that an unknown ECU is classified as one of the valid ECUs.
Fig. \ref{Top - Overall Success Rate, Bottom - Alien Detection Rate} shows both FN and FP rates corresponding to varying threshold.
When we defined a higher threshold, FN rates went up, while FP rates went down.
Even though SVM did not output the best performance among three different classification algorithms in Subsection \ref{Basic Identification}, SVM can perfectly distinguish between known ECUs and unknown ECUs by utilizing a threshold of between $-2.6\times10^{15}$ and $-2.4\times10^{9}$ at which EER (Equal Error Rate) is 0.
For NN, EER is 0.38 at a threshold of 0.9999, meaning that it is not available to identify an alien.
As a result, we conclude our method outputs the most acceptable result when either SVM with linear kernel function or BDT with 100 classifiers is used.

Both CAN2USB-interface and ELM327 are typical devices that can be connected to the OBD II port and adversaries can purchase those devices quite inexpensively \cite{ELM327,OBD2USB}. Accordingly, we assume that those devices are likely candidates for launching attacks. Using those devices, we sampled more than 1,000 observations or signals per device.

\section{Related Works}
Fingerprinting is a well-established biometrics techniques for identifying human beings \cite{Information_fusion_in_biometrics,capacity_and_examples_of_template-protecting_biometric_authentication_system}. Beginning in the 20th century, the concept of fingerprinting has been used to identify devices as well.
In particular, World War II made it both important and popular to try to identify radar, radios, and other wireless communications \cite{Most_secret_war}.
Many researchers have continued to study identification of wireless signals right up to the present \cite{Characteristics_of_radio_tranmitter_fingerprints,Detection_of_radio,barbeau2003intrusion,hall2003detection,ferrell1991method,mensa2003radar,frederick1995cellular}.
These methods are usually based upon minor variations in analog hardware in transmitters, utilizing the individual characteristics of the transmitters \cite{brik2008wireless}.
Fundamentally, the approach to identifying devices according to their unique characteristic recognizes that physical device are all different in the way they produce signals, even when they are produced by the same manufacture.

Some previous works has identified network devices by using the clock skew of the devices. On a network such as the Internet closck skew describes the difference in time shown by the clocks at different nodes on the network.
Accordingly, it is possible to distinguish devices through TCP and ICMP timestamps \cite{kohno2005remote}.
Previous research has studied not only hardware-based fingerprinting, but also software-based fingerprinting. Software-based fingerprinting uses patterns traceable to a user.
For example, by analyzing month-logs, users can be tracked or identified \cite{yen2012host}.
Browser history can also be used for tracking or identifying users \cite{olejnik2012johnny}.
Software-based fingerprinting uses a uniqueness of users' patterns such as current configuration of the system (e.g., screen resolution, or list of fonts installed in a system) \cite{acar2013fpdetective}.

Recently, since various smart devices including smartphones are being developed, there has been a substantial amount of work done on smart device fingerprinting.
In order to identify smart devices, researchers use smart device sensor.
Camera sensor identification is one of the main methods for identifying smart devices \cite{lukas2006digital,li2010source}.
Based on the sensor's pattern noise, it is possible to distinguish images for the same scene.
The images include unique differences because camera sensors are all different from one another.
The embedded acoustic components of smartphones have also been used to identify smartphones \cite{das2014you,bojinov2014mobile}.
Moreover, different microphones convert the same sound source into different electronic signals, or conversely loudspearkers convert the same audio signals into different sounds in the air.
By using the unique characteristics of each device, it is possible to identify microphones and loudspeakers when analyzed separately or together.
Another approach to identifying sensors in smartphones is to use accelerometer \cite{Accelprint,bojinov2014mobile}.
Accelerometers measure 3-axis acceleration and regulate, for example, smart phone screen rotation.
The measured values are a little bit different in each accelerometer.
Compared to other methods mentioned, which compare devices emitting the same sound or image, accelerometers use gravitational acceleration (i.e., $9.8m/s^2$) that is always detectable.
Accelerometers show that we do not have to be using the artificial same source to be identified by sensor devices.

Similar to our work, there are several researchers who have identified devices or transmitters in wired communication. Gerdes et al. proposed a method for identification of wired Ethernet devices \cite{Wired_physical_layer}.
They used the matched filter (known as a North filter) to compute the correlation between a target signal and a reference signal and thereby identify the devices. Murvay et al. proposed a method for source identification using signal characteristics in CANs (Controller Area Networks) \cite{Cosic}.
Their work was the first approach to identifying ECUs using inimitable characteristic of signals. However, the success rate of their method is not sufficient to effectively identify attackers in a real environment.
Furthermore, their experimental model did not sufficiently simulate critical functions of the in-vehicle CAN network.

\section{Limitation and Future Works}
In this section, we discuss limitations of our method and future work to improve the method. Our method uses the extended frame format instead of base frame format. Even though this frame format is part of the standard CAN environment, current in-vehicle networks currently function solely with the base frame format. Thus, our method can be used without changing ECU hardware, but ECU firmware would need to be updated in order to use the extended frame format.  This firmware update would involve new programming in new vehicles or reprogramming in older cars, but again,  replacement of existing ECU hardware is not required. Because our method considers that one ECU belongs to one class, we need a fixed bit string for one ECU. However, if it is possible to get enough long fixed bit string form base frame format not extended frame format, we do not have to update existing firmware. As a result, we plan to do future work to develop a method without using the extended frame format.

\section{Conclusion}
We have presented our new method to identify ECUs using inimitable characteristics of signals emitted from different ECUs. Since our method only requires installation of a monitoring unit to the in-vehicle CAN network, without any change in ECU hardware, our method can be directly and transparently applied into the current system. In addition, our method adequately simulates the CAN standard and is compatible with it. Thus, we have minimized the required cost of applying this security method to the existing in-vehicle CAN network. We identified the limitations in Mulray's work \cite{Cosic}, improving it to be effective in a real environment. Our method is more than twice as accurate as the previous research by Murvay et al. \cite{Cosic}, as reflected in the low false positive rate. Moreover, we tested  our method under the same environment as the actual in-vehicle CAN network. As a result, we conclude our proposed method is a feasible and effective approach that can realistically be applied into in-vehicle CAN network.

\section{ACKNOWLEDGMENT}
This work was supported by Samsung Research Funding Center of Samsung Electronics under Project Number SRFC-TB1403-02.

\bibliographystyle{IEEEtranS}
\bibliography{IEEEabrv,bibliography}

\end{document}